# Four-Wave Mixing in Long Wavelength III-Nitride QD-SOAs

Amin H. Al-Khursan[1,*], A. Subhi[2] and H. I. Abood[3]

[1]Nassiriya Nanotechnology Research Laboratory (NNRL), Science College, Thi-Qar University, Nassiriya, Iraq.

[2]Physics Department, Science College, Al-Muthana University, Samawah, Iraq.

[3]Physics Department, Science College, Babylon University, Hillah, Iraq.

* Corresponding author: ameen_2all@yahoo.com

PACS number(s): 78.67.Hc, 78.67.De, 77.84 Bw.

**Abstract:** Four wave mixing analysis is stated for quantum dot semiconductor optical amplifiers (QD SOAs) using the propagation equations (including nonlinear propagation contribution) coupled with the QD rate equations under the saturation assumption. Long wavelength III-nitride InN and AlInN QD SOAs are simulated. Asymmetric behavior due to linewidth enhancement factor is assigned. P-doping increases efficiency. Lossless efficiency for InAlN QDs for longer radii is obtained. Carrier heating is shown to have a considerable effect and a detuning dependence is expected at most cases. InN QD SOAs shown to have higher efficiency.

**Keywords: four-wave mixing, quantum dot, semiconductor optical amplifier.**

## 1. Introduction

Four-wave mixing (FWM) is one of the best ways for coherently converting optical signals from one frequency to another, reducing the network channels and preventing wavelength blocking, thus developing the optical communications networks the mature of this new century. Using quantum dots (QDs) in the active region of the semiconductor optical amplifier (SOA) gets better attention in recent years due to the demonstrated features, mainly: high saturation power and broad gain bandwidth results in high efficiency conversion. QD-SOAs also attractive in eliminating destructive interference and obtaining high efficiency required for signal to noise ratio, needed for data transmission, the problem facing long wavelength conversion when bulk and quantum well SOAs are used [1,2].

FWM in QDs is related to three mechanisms: spectral hole burning (SHB), carrier density pulsation (CDP) and carrier heating (CH) which has small contribution at some QD semiconductors. Although the three mechanisms are the same as other bulk and quantum well structures but their origin here is different due to QD inhomogeneity which results from size nonuniformity and manufacture imperfections. SHB occurs as the carriers in the ground states (GS) only depletes by the pump field creating a spectral hole. Then, other carriers from excited states (ES) in the same QD relax to it. Carrier relaxation time from QD GS to ES is typically 100-250fs [3]. Due to these slower relaxation mechanisms, efficient FWM in QDs is expected where deeper spectral holes are related to these relaxations. In the CDP, the entire gain curve is reduced due to the decrease of the total carrier density in the QDs [1]. The third mechanism is the CH where lowest energy carriers removal from GS are due to stimulated emission and increasing energy of free carriers by photon absorption. Thus energy and temperature of carrier distribution increased while the lattice temperature remains unchanged. These hot carriers are then cool down by carrier-phonon collisions. Either of density matrix theory or propagation wave equations is used to derive FWM models [1-5]. No one of these theories discusses the nonlinear wave propagation effect in QD SOAs. Following the way of Mecozzi et al. [6] we derive an analytical model describing FWM in QD SOAs using the propagation equations (including nonlinear propagation contribution) coupled with the QD rate equations using the assumption relevant for the applications in which the output probe signal higher than the output of the conjugate. ES is included in the analysis in addition to GS of QDs. The study covers III-nitride QD-SOAs working at the mid-infrared region where they get, now, a considerable importance in researches due to their application in high-power high speed and telecommunications [7].

## 2. Four Wave Mixing Theory

### 2.1 Rate Equations System

The rate equations used to describe carrier density dynamics in the QD-SOA is the following system

$$\frac{dN_{WL}}{dt}=\frac{J}{eL_W}-\frac{N_{WL}}{\tau_{wr}}-\frac{N_{WL}}{\tau_{we}}+\frac{N_{ES}}{\tau_{ew}} \qquad (1a)$$

$$\frac{dN_{ES}}{dt}=\frac{N_{WL}}{\tau_{we}}+\frac{N_{GS}}{\tau_{ge}}-\frac{N_{ES}}{\tau_{eg}}-\frac{N_{ES}}{\tau_{ew}}-\frac{N_{ES}}{\tau_{r}} \qquad (1b)$$

$$\frac{dN_{GS}}{dt}=\frac{N_{ES}}{\tau_{eg}}-\frac{N_{GS}}{\tau_{r}}-\frac{N_{GS}}{\tau_{ge}}-\Gamma g_{GS}|E|^2 \qquad (1c)$$

The carrier dynamics processes of the above system are schematically represented in Fig. 1. Electrons are injected with a current density $J$ into the $In_xAl_{1-x}N$ wetting layer (WL) with a thickness $L_W$ (the effect of clad layer is neglected) then captured at a relatively slow-rate ($\tau_{we}$) into the InN (or $In_xAl_{1-x}N$) QD ES, which works as a common reservoir. an escape components from ES to WL at a rate ($\tau_{ew}$) and from GS to ES at a rate ($\tau_{ge}$) is also included. The carriers are then captured in the GS at a rate ($\tau_{eg}$). The spontaneous radiative life time components are also contributes in these dynamics at rates ($\tau_{wr}$) in the WL and ($\tau_r$) in the QD GS and ES. In the system of Eqs. (1), $e$ is the electron charge, $\Gamma$ is the optical confinement factor, $g_{GS}$ is the gain assumed here due to QD GS transitions only and the contribution of the ES is neglected. This assumption coincides with many recent literatures in this field. To include the propagation effects [8], we began from the propagation wave equation in the QD active region which can be written as follows

$$\frac{dE(z,t)}{dz} = \left(\frac{1}{2}g + ik_{nl}\right)E(z,t) \qquad (2)$$

$E(z,t)$ is the electric field amplitude in the QD-SOA and can be written as

$$E(z,t) = A_j(z)\exp(iw_j t) \qquad (3)$$

where *j=0, 1, 2* for pump, probe, and conjugate waves, respectively, *i* is the imaginary number while the slowly varying envelope $A_j(z)$ is defined as

$$A_j(z) = E_j(z)\exp(ik_j z) \qquad (4)$$

The frequency $w_j$ is defined as

$w_j = w_0 - l\Omega$, $l = 0,-1,+1....$ For *j=0, 1, 2*, respectively.

Then the electric field amplitude becomes

$$\begin{aligned} E(z,t) &= E_1\exp(ik_1 z - i(w_0+\Omega)t) + E_0\exp(ik_0 z - iw_0 t) \\ &\quad + E_2\exp(ik_2 z - i(w_0-\Omega)t) \end{aligned} \qquad (5)$$

The three beams in which the electrical field has been decomposed are the pump $E_0$ at frequency $w_0$, the probe $E_1$ at frequency $w_0+\Omega$ and the conjugate $E_2$ at frequency $w_0 - \Omega$, see Fig. 2. The total gain in the QD SOA including spectral-hole burning and carrier heating can be written as [6]

$$g_{GS} = a(N_{GS} - N_0)\left[1 - \varepsilon_{sh}\int_0^{\infty} h_{sh}(t')\left|E(t-t')\right|^2 dt'\right] - g_o\varepsilon_{ch}\int_0^{\infty} h_{ch}(t')\left|E(t-t')\right|^2 dt' \qquad (6)$$

, for the nonlinear contribution to the wave number we have

$$k_{nl} = \frac{1}{2}\left[\alpha a(N_{GS} - N_0) - g_o\varepsilon_{ch}\beta\int_0^{\infty} h_{ch}(t')\left|E(t-t')\right|^2 dt'\right] \qquad (7)$$

This contribution recognizes the model from other FWM models. Here, $a$ is the differential gain factor, $N_{GS}$ is the carrier density in the GS, $N_o$ is the carrier density at transparency, $\alpha$ is the linewidth enhancement factor. The linear response functions of the gain to a change of carrier temperature and distribution of intraband population are $h_{ch}(t)$ and $h_{sh}(t)$, respectively. The parameters $\varepsilon_{ch}$ and $\varepsilon_{sh}$ are the strength of the corresponding nonlinear processes. The normalization gain $g_o$ has been introduced to have the same units of $\varepsilon_{ch}$ and $\varepsilon_{sh}$. The effect of the change of the electron temperature on the index is introduced through the parameter $\beta$. The response functions are normalized to one, i.e.

$$\int_0^\infty h_{ch}(t')dt' = \int_0^\infty h_{sh}(t')dt' = 1 \qquad (8)$$

The Fourier transform of the response functions are defined as

$$h(\Omega) = \int_0^\infty h(t')\exp(i\Omega t')dt' \qquad (9)$$

It can be expressed analytically for both spectral hole-burning and carrier heating as follows [9]

$$h_{ch}(\Omega) = \frac{1}{(1-i\Omega\tau_1)(1-i\Omega\tau_2)} \qquad (10)$$

$$h_{sh}(\Omega) = \frac{1}{(1-i\Omega\tau_2)} \qquad (11)$$

where $\tau_1$ is the characteristic time of the carrier-LO phonon scattering, responsible for the cooling of the carrier distribution to the lattice temperature and $\tau_2$ is the characteristic time of the carrier-carrier scattering responsible for the filling of the hole burned by the field in the intraband carrier distribution and for the heating of the electron and hole gas. The carrier densities due to the beating between pump and probe fields can be beat in the following form

$$N_{WL} = \overline{N}_{WL} + \Delta N_{WL}\exp(-i\Omega t) + \Delta N_{WL}^{*}\exp(i\Omega t) \qquad (12a)$$

$$N_{ES} = \overline{N}_{ES} + \Delta N_{ES}\exp(-i\Omega t) + \Delta N_{ES}^{*}\exp(i\Omega t) \qquad (12b)$$

$$N_{GS} = \overline{N}_{GS} + \Delta N_{GS} \exp(-i\Omega t) + \Delta N_{GS}^{*} \exp(i\Omega t) \qquad (12c)$$

We enter Eq. (12a) into Eq. (1a), to obtain three equations for the time-independent term and for the coefficients of $\exp(\pm i\Omega t)$, then we neglect the coefficients of $\exp(\pm i\Omega t)$, to get, for steady-state WL carrier density

$$\overline{N}_{WL} = \left[ \frac{J}{eL_W} + \frac{\overline{N}_{ES}}{\tau_{ew}} \right] / \left[ \frac{1}{\tau_{wr}} + \frac{1}{\tau_{we}} \right] \qquad \mathbf{(13)}$$

By the same way, entering Eqs. (12b) and (13) into Eq. (1b), to get for steady-state ES carrier density

$$\overline{N}_{ES} = \left\{ \left[ \left( \frac{1}{\tau_{we}} \frac{J}{eL_W} \right) / \left( \frac{1}{\tau_{wr}} + \frac{1}{\tau_{we}} \right) \right] + \overline{N}_{ES} \frac{1}{\tau_{ge}} \right\} / \left\{ \frac{1}{\tau_{eg}} + \frac{1}{\tau_{ew}} \right.$$

$$\left. + \frac{1}{\tau_r} - \left[ \left( \frac{1}{\tau_{we}} + \frac{1}{\tau_{ew}} \right) / \left( \frac{1}{\tau_{wr}} + \frac{1}{\tau_{we}} \right) \right] \right\} \qquad \mathbf{(14)}$$

Using Eqs. (12c), (14) and (6) after neglecting all the terms proportional to $\varepsilon_{ch}$ and $\varepsilon_{sh}$, in Eq. (1c), neglecting the coefficients of $\exp(\pm i\Omega t)$, to get, for steady-state case

$$\frac{J}{eL_W} \frac{\tau_{wr}}{\tau_{we} + \tau_{wr}} \frac{\tau_{ge}}{\tau_{DR}} - \overline{N}_{GS} \frac{1}{\tau_D} - \Gamma a(\overline{N}_{GS} - \overline{N}_{GSo})S - \Gamma a \Delta N_{GS} (E_0 E_1^{*} + E_2 E_0^{*})$$

$$-\Gamma a \Delta N_{GS}^{*}(E_1 E_0^{*} + E_0 E_2^{*}) = 0 \qquad (15)$$

where

$$\frac{1}{\tau_D} = \frac{1}{\tau_r} + \frac{1}{\tau_{ge}} - \frac{1}{\tau_{DR}} \qquad (16)$$

$$\tau_{DR} = \tau_{eg}\tau_{ge}\left\{\frac{1}{\tau_{eg}} + \frac{1}{\tau_{ew}} - \frac{1}{\tau_r} - \left(1/\left[\tau_{we}\tau_{ew}\left(\frac{1}{\tau_{wr}} + \frac{1}{\tau_{we}}\right)\right]\right)\right\} \qquad (17)$$

$$S = |E_0|^2 + |E_1|^2 + |E_2|^2 \qquad (18)$$

Neglecting the terms proportional to $\Delta N_{GS}$ $\Delta N_{GS}^{*}$ in Eq. (15) to get

$$\Gamma a(\overline{N}_{GS} - \overline{N}_{GSo}) = \frac{\Gamma \overline{g}}{1 + S/P_{sat}} \qquad (19)$$

where

$$\overline{g} = \frac{aJ}{eL_W}\frac{\tau_{wr}\tau_D}{\tau_{we} + \tau_{wr}}\frac{\tau_{ge}}{\tau_{DR}} - a\overline{N}_{GSo} \qquad (20)$$

$$P_{sat} = \frac{1}{\Gamma a \tau_D} \qquad (21)$$

$\overline{g}$ is the unsaturated gain coefficient, and $P_{sat}$ is the saturation power, introducing $p$ as a new parameter defined as

$$p(z,t) = \frac{|E(z,t)|^2}{P_{sat}} \qquad (22)$$

By using Eq. (5), we get

$$p(z,t) = p_S(z) + p_{12}(z)\exp(-i\Omega t) + p_{21}(z)\exp(i\Omega t) + ... \qquad (23)$$

where

$$p_S = S / P_{sat} \qquad (24)$$

from Eq. (18), we get

$$p_S = p_0 + p_1 + p_2 \qquad (25a)$$

where

$$p_0 = |E_0|^2 / P_{sat} \qquad (25b)$$

$$p_1 = |E_1|^2 / P_{sat} \qquad (25c)$$

$$p_2 = |E_2|^2 / P_{sat} \qquad (25d)$$

and

$$p_{12} = (E_1 E_0^* + E_2^* E_0) / P_{sat} \qquad (26)$$

$$p_{21} = (E_2 E_0^* + E_1^* E_0) / P_{sat} \qquad (27)$$

Using Eqs. (12c), (14) and (6) after neglecting all the terms proportional to $\varepsilon$, in Eq. (1c), and taking the terms proportional $\exp(\pm i\Omega t)$ to get

$$\Gamma a \Delta N_{GS} = -\frac{\Gamma a(\overline{N}_{GS} - \overline{N}_{GSo}) p_{12}}{(1 + p_S - i\Omega\tau_D)} \qquad (28)$$

The propagation equations for the pump, probe and conjugate, can be obtained by substituting Eqs. (5), (6), (7) and (12c) into Eq. (2)

$$\frac{dp_0}{dz} = \frac{1}{4}\left\{\gamma_{sc}^2 - \frac{\Gamma\overline{g}}{1+p_S}\left[2\gamma_{sc} - (1+\alpha^2)\frac{\Gamma\overline{g}}{1+p_S}\right]\right\} p_0 \qquad (29)$$

Note that $\gamma_{sc}$ is a loss coefficient introduced to include the scattering loss of the waveguide. Also, for probe and conjugate field propagation equations they can be obtained by considering all the terms oscillating at frequency $w_0 + \Omega$ and $w_0 - \Omega$ respectively, as follows

$$\frac{dp_{1,2}}{dz} = \frac{\left|\Xi_{1,2}\right|^2}{P_{sat}} \qquad (30)$$

where

$$\Xi_{1,2} = \frac{1}{2}\left\{-\gamma_{SC} + (1-i\alpha)\frac{\Gamma\overline{g}}{1+p_S}\left[1 - \frac{p_0}{1+p_S \pm i\Omega\tau_D}\right]\right\} E_{1,2} - \frac{1}{2}(1-i\alpha)\frac{E_0^2 E_{2,1}^*}{P_{sat}}\frac{1}{1+p_S \pm i\Omega\tau_D}\frac{\Gamma\overline{g}}{1+p_S}$$
$$-\frac{1}{2}(1-i\beta)\frac{E_0^2 E_{2,1}^*}{P_{sat}}\Gamma g_o \varepsilon_{ch} P_{sat} h_{ch}(\mp\Omega) - \frac{1}{2}\frac{E_0^2 E_{2,1}^*}{P_{sat}}\frac{\Gamma\overline{g}}{1+p_S}\varepsilon_{sh} P_{sat} h_{sh}(\mp\Omega) \qquad (31)$$

the upper sign refers to the probe $E_1$ and the lower to the conjugate $E_2$.

## 2.2 Mode Equations Solution

Three nonlinear equations, Eqs. (29) and (30), are obtained. Their analytical solution can be done under conditions that are often met virtually. The pump and probe are only entered while the conjugate is generated through the FWM process. The basic approximation is that we neglect in Eq. (30) for the probe all the terms proportional to the conjugate, while pump and conjugate are stay with no approximations. This is a good approximation if the intensity of the conjugate is much lower than the intensity of the probe at output. For pump wave one can write

$$\frac{dp_0(z)}{dz} = \left|A_0[p_S(z)]\right|^2 p_0(z) \qquad (32)$$

where

$$A_0[p_S(z)] = \frac{1}{2}\left\{-\gamma_{SC} + (1-i\alpha)\frac{\Gamma\overline{g}}{1+p_S}\right\} \qquad (33)$$

while for the probe

$$\frac{dp_1(z)}{dz} = \left|A_1[p_S(z)]\right|^2 p_1(z) \qquad (34)$$

where

$$A_1[p_S(z)] = \frac{1}{2}\left\{-\gamma_{SC} + (1-i\alpha)\frac{\Gamma\overline{g}}{1+p_S}[1-\frac{p_0}{1+p_S+i\Omega\tau_D}]\right\} \qquad (35)$$

For the conjugate one can write also

$$\frac{dp_2(z)}{dz} = |A_2[p_S(z)]|^2 p_2(z) + p_0^2(z) p_1(z) |B[p_S(z)]|^2 + A_2[p_S(z)] E_2(z) \frac{E_0^{*2}(z) E_1(z)}{P_{sat}^2} B^*[p_S(z)]$$
$$A_2^*[p_S(z)] E_2^*(z) \frac{E_0^2(z) E_1^*(z)}{P_{sat}^2} B[p_S(z)] \qquad (36)$$

where

$$A_2[p_S(z)] = \frac{1}{2}\left\{ -\gamma_{SC} + (1-i\alpha)\frac{\Gamma \overline{g}}{1+p_S}[1 - \frac{p_0}{1+p_S - i\Omega\tau_D}] \right\} \qquad (37)$$
$$B[p_S(z)] = -\frac{1}{2}(1-i\alpha)\frac{E_0^2 E_{2,1}^*}{P_{sat}} \frac{1}{1+p_S \pm i\Omega\tau_D} \frac{\Gamma \overline{g}}{1+p_S} - \frac{1}{2}(1-i\beta)\frac{E_0^2 E_{2,1}^*}{P_{sat}} \Gamma g_o \varepsilon_{ch} P_{sat} h_{ch}(\mp\Omega)$$
$$-\frac{1}{2}\frac{E_0^2 E_{2,1}^*}{P_{sat}} \frac{\Gamma \overline{g}}{1+p_S} \varepsilon_{sh} P_{sat} h_{sh}(\mp\Omega) \qquad (38)$$

One can see that $A_1$ and $A_2$ are also functions of $P_0$, not only for $P_S(z)$. Eqs. (32), (34) and (36) are the approximated set that we are going to solve analytically. Since we have assumed that the conjugate is weak and generated through the FWM process, then the total power $P_s$ is taken as the sum of the pump and probe powers only

$$p_S = p_0 + p_1 \qquad (39)$$

Now we are consider two cases, low and high detuning.

## Low Detuning

When the detuning frequency is much lower than $\Omega_s = \frac{1}{\tau_D}$. The saturation of the amplifier is set by pump and probe together. From Eqs. (32) and (34) one gets

$$\frac{dp_0}{dz} = \frac{1}{2}\left\{-\gamma_{sc} + \frac{\Gamma\bar{g}}{1+p_S}\right\}p_0 \qquad (40)$$

and

$$\frac{dp_1}{dz} = \frac{1}{2}\left\{-\gamma_{sc} + \frac{\Gamma\bar{g}}{1+p_S}\right\}p_1 \qquad (41)$$

Now using Eqs. (39)-(41), one gets

$$\frac{dp_S}{dz} = \frac{1}{2}\left\{-\gamma_{sc} + \frac{\Gamma\bar{g}}{1+p_S}\right\}p_S \qquad (42)$$

This equation is in a closed form and can be solved analytically by variables separation when the pump and probe power are given at input, see Appendix. The final result is

$$p_S(0) = \left(\frac{1-\zeta}{\zeta}\right)\frac{1-\chi}{G-\chi} \qquad (43)$$

where

$$\chi = \left( \frac{G}{G_o} \right)^{\zeta} \qquad (44)$$

$$G_o = \exp[(\Gamma \bar{g} - \gamma_{sc})z] \qquad (45)$$

$$G = \frac{p_S(z)}{p_S(0)} \qquad (46)$$

$$\zeta = \frac{\gamma_{sc}}{\Gamma \bar{\bar{g}}} \qquad (47)$$

The parameter $G_o$ is the unsaturated (single pass) gain of the amplifier, and $G$ is the saturated gain at position $z$. The total gain of the amplifier corresponds to $z = L$, the total length of the device. Given the values of $G_o$ and $p_S(0)$, the value of $G$ can be obtained by numerical solution for Eq. (43). From the relation between $P_0$ and $P_1$, using Eqs. (40) and (41) one can be find

$$\frac{d}{dz}\frac{p_0(z)}{p_1(z)} = \frac{\left[ -\gamma_{sc} + \dfrac{\Gamma \bar{g}}{1+p_S} \right] p_0(z)}{\left[ -\gamma_{sc} + \dfrac{\Gamma \bar{g}}{1+p_S} \right] p_1(0)} = \frac{p_0(z)}{p_1(0)} = \sigma' \qquad (48)$$

also

$$\frac{d}{dz}\frac{p_0(z)}{p_1(z)} = \frac{\sigma'}{\sigma' + 1} = \sigma$$

$$\sigma = \frac{p_0(z)}{p_S(z)}$$

or

$$p_0(z) = \sigma \quad p_S(z) \qquad (49)$$

$\sigma$ can be calculated at $z = 0$

$$p_0(0) = \sigma \quad p_S(0)$$

Return back to Eqs. (32), (34) and (36), they can be solved after some mathematical manipulation, to get

$$p_0(z) = p_0(0)\exp\left[\int_0^z \left|A_0[p_S(z')]\right|^2 dz'\right] \qquad (50)$$

$$p_1(z) = p_1(0)\exp\left[\int_0^z \left|A_1[p_S(z')]\right|^2 dz'\right] \qquad (51)$$

and

$$p_2(z) = \exp\left[\int_0^z \left|A_2[p_S(z')]\right|^2 dz'\right]\int_0^z p_0^2(z')p_1(z')\left|B[p_S(z')]\right|^2 dz' \exp\left[-\int_0^{z'} \left|A_2[p_S(z'')]\right|^2 dz''\right] \qquad (52)$$

Now we substituted Eqs. (50) and (51) into Eq. (52), we get

$$p_2(z) = \exp\left[\int_0^z \left|A_2[p_S(z')]\right|^2 dz'\right]\int_0^z p_0^2(0)p_1(0)\left|B[p_S(z')]\right|^2 dz'$$

$$.\exp\left[\int_0^z \left\{2\left|A_0[p_S(z')]\right|^2 + \left|A_1[p_S(z')]\right|^2 - \left|A_2[p_S(z')]\right|^2\right\}dz'\right] \qquad (53)$$

Using Eq. (53) and after some mathematical manipulation, one can obtains

$$p_2(z) = p_1(0)|\xi|^2 G \qquad (54)$$

and

$$p_1(z) = p_1(0) G \exp[\mathrm{Re}(-(1-i\alpha)\sigma I_{cd}(z,-\Omega))] \qquad (55)$$

where

$$\xi = -\frac{1-i\alpha}{\alpha}\exp\left[-\frac{1}{2}\sigma I_{cd}(z,\Omega)\right]\sin\left[\frac{\alpha}{2}\sigma I_{cd}(z,\Omega)\right] - \frac{1}{2}(1-i\beta)\frac{g_o}{g}\varepsilon_{ch} h_{ch}(\Omega)\sigma I_{ch}(z)$$
$$-\frac{1}{2}\varepsilon_{sh} h_{sh}(\Omega)\sigma I_{sh}(z) \qquad (56)$$

$$I_{sh}(z) = Ln\left(\frac{G_o}{G}\right) \qquad (57)$$

$$I_{cd}(z,\Omega) = \frac{1}{1-i\Omega\zeta\tau_D}\left[Ln\left(\frac{1+p_S(0)G-i\Omega\tau_D}{1+p_S(0)-i\Omega\tau_D}\right)+\zeta \quad I_{sh}(z)\right] \qquad (58)$$

$$I_{ch}(z) = -\frac{1}{\zeta}\left[p_S(0)(G-1)-I_{sh}(z)\right] \qquad (59)$$

## High Detuning

When detuning is much larger than $\Omega_s$, the high-frequency detuning can be found simply by substitution $\gamma_{sc}=0$ in Eqs. (42)-(59), the final result is

$$p_S(0) = \left(\frac{1}{G-1}\right) I_{sh}(z) \qquad (60)$$

$$p_2(z) = p_1(0)|\xi|^2 G \qquad (61)$$

$$p_1(z) = p_1(0)G \qquad (62)$$

$$\xi = -\frac{1}{2}\sigma\left\{(1-i\alpha)I_{cd}(z,\Omega) + (1-i\beta)\frac{g_o}{\bar{g}}\varepsilon_{ch}h_{ch}(\Omega)I_{ch}(z) + \varepsilon_{sh}h_{sh}(\Omega)I_{sh}(z)\right\} \qquad (63)$$

$$I_{sh}(z) = \ln\left(\frac{G_o}{G}\right) \qquad (64)$$

$$I_{cd}(z,\Omega) = \frac{I_{sh}(z)}{1-i\Omega\tau_D} \qquad (65)$$

$$I_{ch}(z) = I_{sh}(z) + \frac{G+1}{2(G-1)}(I_{sh}(z))^2 \qquad (66)$$

### 2.3 Normalized Conjugate Output and the Efficiency

Two relevant quantities in FWM to be calculated, they are

1. The output power of the conjugate normalized to the output power of the probe

$$\rho = \frac{p_2(z)}{p_1(z)} \qquad (67)$$

At low detuning it becomes

$$\rho = |\xi|^2 \exp\left[\mathrm{Re}\left(\frac{1}{2}(1-i\alpha)\sigma I_{cd}(z,-\Omega)\right)\right] \qquad (68)$$

While at high detuning

$$\rho = |\xi|^2 \qquad (69)$$

2. FWM efficiency, defined as the ratio between the output power of the conjugate and the input power of the probe

$$\eta = \frac{p_2(z)}{p_1(0)} \qquad (70)$$

At low detuning it is written as

$$\eta = G|\xi|^2 \qquad (71)$$

While at high detuning it becomes

$$\eta = G|\xi|^2 \qquad (72)$$

### 3. Specification of the Structures

Shapes of QDs studied here are assumed to be in the form of quantum discs. Two classes of structures are specified in Table 1. In class A we change the WL composition while in class B the In-mole fraction in the QD layer is changed. The quantum discs in these structures have a *2*nm height and *14*nm radius. For each class we chose one of the structures then changing disc height from (*1-2.5*nm) then radius from (*12-15*nm). This is to examine the effect of QD, WL composition and to study quantum size effect on the QD SOA. In addition to this, it is also to cover the possible spectral ranges in the III-nitride structures. Energy levels for the QDs are calculated using the parabolic band quantum-disc model [10,11]. The accuracy of the quantum disc model is checked by the comparison with experimental data or with numerical methods. Strain between QDs and WL is included in the energy levels calculation and is assumed to have a uniform distribution. Thus, it is adequate to calculate energy levels without tedious or time consumption. WL is assumed to be in the form of a quantum well with (*5* nm) thickness and their energy levels are also calculated using parabolic band model. In the calculation of quasi-Fermi energy, in addition to QD levels, only the ground state of the WL conduction and valence bands are included. Gain is then calculated assuming inhomogeneously broadened dots and using a Gaussian lineshape function. This is described well in Ref. 12. Doping for these structures is also examined. Tables 2-3 specify the maximum gain and peak wavelengths for QD structures studied. These data are then used in the FWM calculations of QD SOAs. For each structure, it is assumed that the input signal have a wavelength of the GS transition of the QD. Other parameters used are listed in Table 4.

## 4. Calculations and Discussion

Figure 3(a) and (b) shows the conversion efficiency for InN QD-SOA. Since the linewidth enhancement factor is included in our analysis, thus asymmetric positive and negative detuning branches are obtained. Although zero linewidth enhancement factor (LEF) can be obtained at some points of operation, but the injected carriers in the barrier and WL shown to give large variations in the QD LEF. This is known in the literature as the plasma effect [13]. In Fig. 3 (b), increasing input powers for the pump $(p_0)$ and probe $(p_1)$ and fixing its increment at the ratio $(p_1 = \frac{P_0}{10})$, increases the conversion efficiency. A plateau behavior is seen for negative detuning at (*10-50* GHz), then it falling off. To find its origin let's examine the contributions from FWM components. This can be seen in Fig. 4 where CH is shown to gives higher contribution than SHB which can be attributed to slow thermal relaxation rate occurs in the QDs. Although this contradicted with most other QD FWM theories [4,13] which considers SHB as the origin of  main contribution in FWM especially at high detuning and high speed, as our case here, but it is with the conclusion from very recent theory which also compared with experiment [3]. This slow relaxation can be seen from WL into the QD excited states. These carriers relaxed are depleted by stimulated emission. Fig. 5 shows the conversion efficiency as a function of their gain values for both p-doped and undoped cases. Increasing gain increases efficiency. The effect of p-doping in increasing efficiency can be seen. Higher gain increase FWM power and then its efficiency. From this figure the leading effect of gain is obvious. But one can expect that at higher gain one can get plateau behavior due to impact of gain saturation. If one considers the two curves of p- and un-doped in the figure continues each other, since p-doping can be considered as carrier increment only, the plateau behavior appears in this figure at higher gain values. Fig. 6 (a) and (b) shows the Efficiency at different dot heights and radii for $In_{0.96}Al_{0.04}N$ QD SOA. Our calculations show that the efficiency reduces with increasing size (height or radius). Carrier relaxation, escape times and number of excited states are known to affect FWM. Since the times are taken constant here as we would like to approach they to experimental values of near structures. Thus here the number of states is change here with changing size. By comparing their subbands number, we show that while the number of QDs is not so important in affecting FWM in shallow (small height) and narrow (small radius) QDs, it appears to be important for deeper QDs. When we check the energy difference between QD states for each size studied we find that shallow (or narrow) dots have large energy difference between states. For QDs with large energy difference between states its carrier escape rate from GS is small then the induced change in the carrier occupation probability in the GS is large in comparison with WL and GS. According to this small QDs have higher FWM efficiency than deeper ones. Note the higher efficiency of p-doped QD structures when the dot height is changed, and especially the saturation behavior of efficiency for the p-doped structures. The last behavior cannot be attributed to the energy difference between states only, it also relates to the number of excited states. A near zero dB behavior is obtained at longer radii InAlN dots. The conversion efficiency is defined as the ratio of the input signal to the output conjugate, so a complete conversion (lossless) is obtained at zero dB. FWM efficiency as a function of the time related to CH $(\tau_1)$ is shown in Fig. 7. At some short times $(\tau_1)$ the efficiency peaks, while neglecting CH reduces the efficiency but it is still high enough. This can also be seen from Fig. 4 where the SHB contribution is also high although the dominance of CH. Detuning dependent efficiency is shown till ~30ps then it becomes independent. We also study the efficiency as a function of the time related to SHB $(\tau_2)$ (not seen here). Approximately the same behavior is obtained. Qasaimeh [4] found that SHB time constant have detuning dependency. He relates this to carrier relaxation and escape times. Fig. 8 (a) shows efficiency for InN QD structure with the carrier relaxation time from ES to GS $(\tau_{eg})$ as a parameter. Efficiency reduces at shorter relaxation to GS. This time constants have detuning dependency. At detuning frequencies higher than 200GHz shorter and longer times have the same behavior. Fig. 8 (b) shows the effect of recombination time $(\tau_r)$ on the efficiency for InAlN QD SOA. While a plateau behavior is shown at longer times till ~ 200GHz detuning, it is shown to raise the efficiency at faster recombination. Also a near (0 dB) behavior is seen for very long recombination time (40ns). Here, also a lossless conversion is obtained for InAlN QD SOA. Fig. 9 (a) and (b) shows the conversion of FWM efficiency for InN and InAlN QD-SOAs, respectively. In Fig. 9 (a) increasing In-mole fraction in the WL increases efficiency till x=0.88 after this mole fraction its efficiency is reduced. This behavior can be attributed directly to the arrangement of their gain values, see Table 2. Fig. 9 (b) shows the efficiency where it increases when the In-mole fraction in the QD increases. Curves arrangement can be attributed also to gain values in Table 2. The bandgap decreases with increasing In-mole fraction. At the same carrier density, we obtain higher gain and then efficiency till some value, then for smaller bandgap structures (x=0.92 and 0.96) transition from excited states must appear thus GS gain is reduced, and then its efficiency, thus it is important to chose the wavelength adequate for the structure used in FWM application. Since the bandgap decrement in the WL with x-values thus one must expect the spillover behavior or increasing of escape from the dot to the WL which must have contribution in efficiency decrement of InN QDs with lower bandgap in the WL. Higher conversion efficiency is obtained for InN QD SOAs compared with InAlN.

## 5. Conclusions

A model analyses four wave mixing in III-nitrides long wavelength quantum dot semiconductor optical amplifiers is stated depending on propagation wave equations coupled (including nonlinear propagation contribution) with quantum dot rate equations under the assumption of saturation. This model shows the asymmetry results from linewidth enhancement factor with carrier heating dominance. InN dots are shown to have higher efficiency than InAlN. Zero efficiency is obtained for some InAlN structures with longer dot radii. Detuning independence at some cases is obtained. Both structures are shown to have enough efficiency for long wavelength applications.

## Appendix

**Solution of Eq. (42)**

From Eq. (42)

$$\frac{dp_S}{dz} = \left\{ -\gamma_{sc} + \frac{\Gamma \bar{g}}{1 + p_S} \right\} p_S \qquad (A.1)$$

$$\frac{(1+p_S)dp_S}{(1-\zeta(1+p_S))p_S} = \Gamma\bar{g}dz \qquad (A.2)$$

Let: $u = 1 + p_S$, $p_S = u - 1$, $dp_S = du$

$$\int \left[ \frac{1/(1-\zeta)}{(1-\zeta u)} + \frac{1/(1-\zeta)}{(u-1)} \right] du = \int \Gamma\bar{g}dz$$

$$\ln(G) + \frac{1}{\zeta}\ln\left( \frac{1-\zeta(1+p_0(0))}{1-\zeta(1+p_0(z))} \right) = \Gamma\bar{g}z - \gamma_{SC}z = \ln(G_o)$$

$$\frac{1}{\zeta}\ln\left( \frac{1-\zeta(1+p_0(0))}{1-\zeta(1+p_0(z))} \right) = \ln(\frac{G_o}{G}) \qquad (A.3)$$

$$\left( \frac{G_o}{G} \right)^{\zeta} = \frac{1-\zeta(1+p_0(0))}{1-\zeta(1+Gp_0(0))} = \frac{1}{\chi}$$

$$p_S(0) = \left( \frac{1-\zeta}{\zeta} \right) \frac{1-\chi}{G-\chi} \qquad (A.4)$$

**Tables**

Table 1. Structures studied in this thesis.

| Structure Type | QD Layer | x-Mole Fraction | QD Height (nm) | QD Radius (nm) |
|---|---|---|---|---|
| A | InN / $In_xAl_{1-x}N$ / $In_{0.25}Al_{0.75}N$ | 0.76-0.94 | 14 | 2 |
| | InN / $In_{0.96}Al_{0.04}N$ / $In_{0.25}Al_{0.75}N$ | | 12, 13, 14, 15 | 2 |
| | InN / $In_{0.96}Al_{0.04}N$ / $In_{0.25}Al_{0.75}N$ | | 14 | 1, 1.5,2,2.5 |
| B | $In_{x.}Al_{1-x}N$/$In_{0.86}Al_{0.14}N$ / $In_{0.25}Al_{0.75}N$ | 0.88-0.96 | 14 | 2 |
| | $In_{0.96}Al_{0.04}N$/$In_{0.86}Al_{0.14}N$ / $In_{0.25}Al_{0.75}N$ | | 12, 13, 14, 15 | 2 |
| | $In_{0.96}Al_{0.04}N$/$In_{0.86}Al_{0.14}N$ / $In_{0.25}Al_{0.75}N$ | | 14 | 1, 1.5,2,2.5 |

Table 2. Results for two type structure with 14 nm radius and 2nm height, at p-doped and the surface electron densities ($n_{2D}$=8.3×$10^{14}$ $m^{-2}$).

| $InN / In_xAl_{1-x}N / In_{0.25}Al_{0.75}N$ | | | |
|---|---|---|---|
| **x-Mole Fraction** | **Peak Material Gain ($cm^{-1}$)** | **Maximum Wavelength (nm)** | **Optical Confinement Factor** |
| 0.96 | 918.303 | 1698 | 0.016 |
| 0.92 | 1722 | 1538 | 0.018 |
| 0.88 | 1803 | 1455 | 0.019 |
| 0.84 | 1644 | 1409 | 0.02 |
| 0.8 | 1591 | 1372 | 0.02 |
| 0.76 | 1427 | 1343 | 0.02 |
| $In_xAl_{1-x}N / In_{0.86}Al_{0.14}N / In_{0.25}Al_{0.75}N$ | | | |
| **x-Mole Fraction** | **Peak Material Gain ($cm^{-1}$)** | **Maximum Wavelength (nm)** | **Optical Confinement Factor** |
| 0.96 | 1339 | 1316 | 0.022 |
| 0.94 | 1072 | 1266 | 0.023 |
| 0.92 | 935.223 | 1201 | 0.025 |
| 0.9 | 802.371 | 1171 | 0.025 |

Table 3. Results for p-doped InN / $In_{088.}Al_{0.12}N$ / $In_{0.25}Al_{0.75}N$ structures for different radii and heights, at the surface electron densities ($n_{2D}=8.3\times10^{14}$ $m^{-2}$).

| Height 2nm | Radius | Peak Material Gain ($cm^{-1}$) | Maximum Wavelength (nm) | Optical Confinement Factor |
|---|---|---|---|---|
| | 12 | 2174 | 1438 | 0.02 |
| | 13 | 1867 | 1460 | 0.019 |
| | 14 | 1803 | 1455 | 0.019 |
| | 15 | 1593 | 1469 | 0.019 |
| **Radius 14nm** | **Height** | **Peak Material Gain ($cm^{-1}$)** | **Maximum Wavelength (nm)** | **Optical Confinement Factor** |
| | 1 | 5719 | 1404 | 0.012 |
| | 1.5 | 2892 | 1425 | 0.016 |
| | 2 | 1803 | 1455 | 0.019 |
| | 2.5 | 1101 | 1523 | 0.021 |

Table 4. List of parameters used in the calculations.

| *Parameter* | *Notation* | *Unit* | *Value* | *Ref.* |
|---|---|---|---|---|
| The carrier relaxation time from WL to ES | $\tau_{we}$ | ps | 3 | 4, 14 |
| The carrier escape time from ES to WL | $\tau_{ew}$ | ns | 1 | 4, 14 |
| The carrier relaxation time from ES to GS | $\tau_{eg}$ | ps | 0.16 | 4, 14 |
| The carrier relaxation time from GS to ES | $\tau_{ge}$ | ps | 1.2 | 4, 14 |
| The spontaneous radiative lifetime in WL | $\tau_{wr}$ | ns | 1 | 4, 14 |
| The spontaneous radiative lifetime in QDs | $\tau_r$ | ns | 0.4 | 4, 14 |
| Carrier heating time ($\tau_{CH}$) | $\tau_1$ | ps | 1 | 15 |
| Hole burning time ($\tau_{SHB}$) | $\tau_2$ | ps | 0.2 | 15 |
| Saturation power | $P_{sat}$ | mW | 10 | |
| Hole burning parameter | $\varepsilon_{sh}$ | $W^{-1}$ | 10 | 9 |
| Carrier heating parameter | $\varepsilon_{ch}$ | $W^{-1}$ | 2.5 | 9 |
| Linear part of Henrys | $\alpha$ | | 4 | 9 |
| Nonlinear part of Henrys | $\beta$ | | 2 | 9 |
| The surface density of QDs | $N_{QD}$ | $m^{-2}$ | $1\times10^{14}$ | 4 |
| Loss coefficient | $\gamma_{sc}$ | $cm^{-1}$ | 3 | 4 |

# Figures

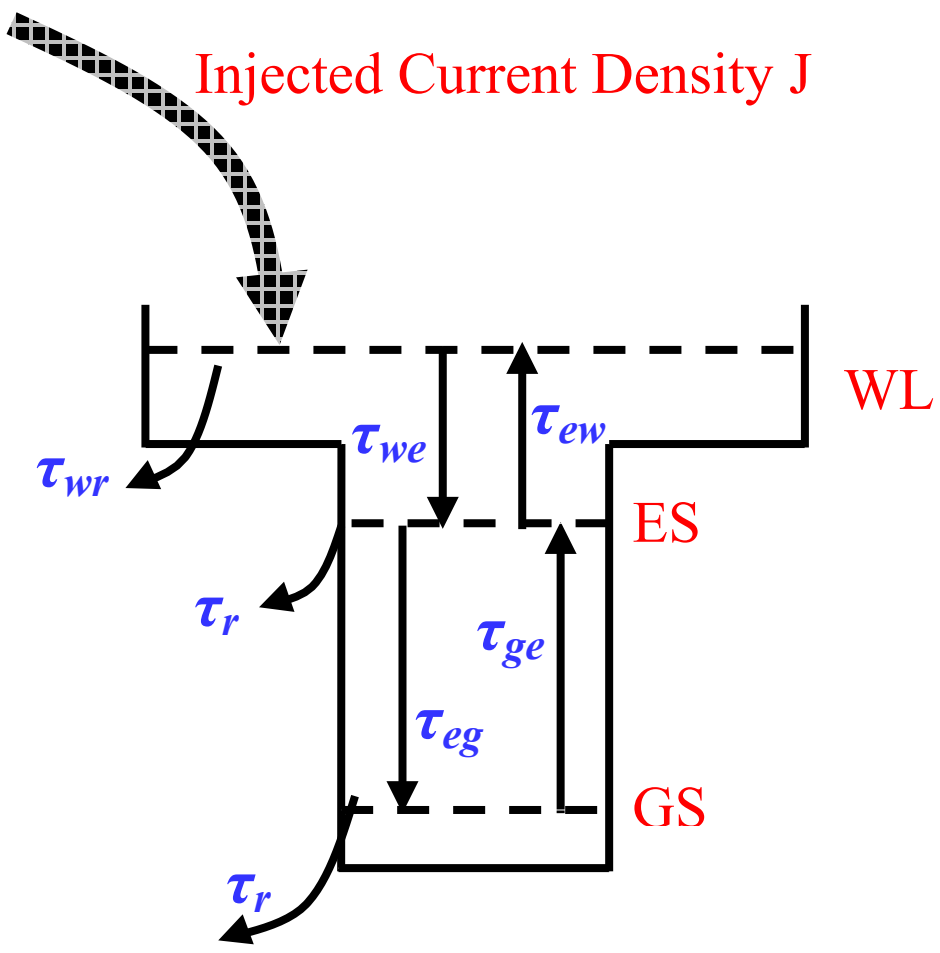

Fig. 1. Energy diagram of a QD portraying the characteristic times for the carrier dynamics model.

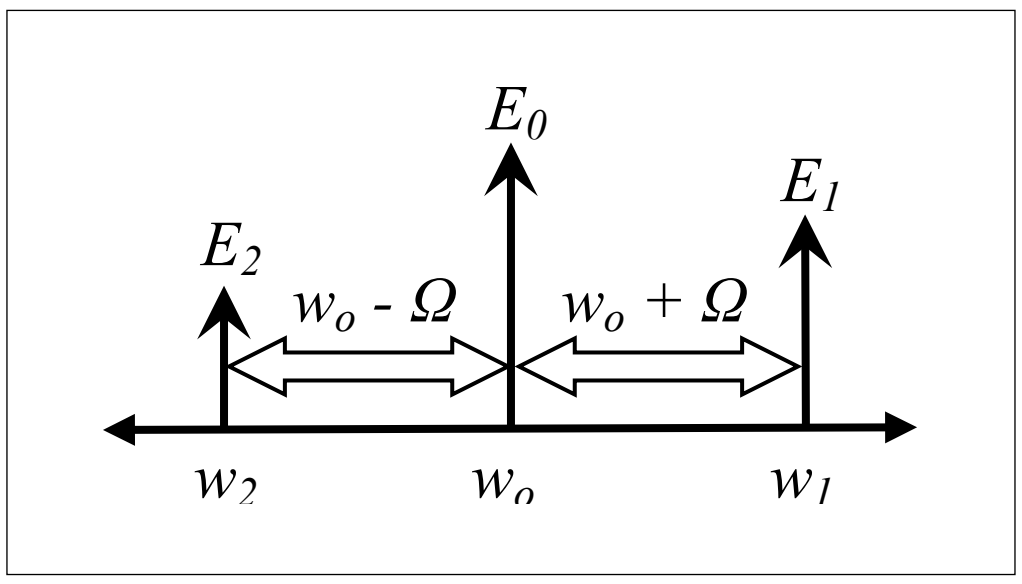

Fig. 2. The physical picture of FWM frequencies.

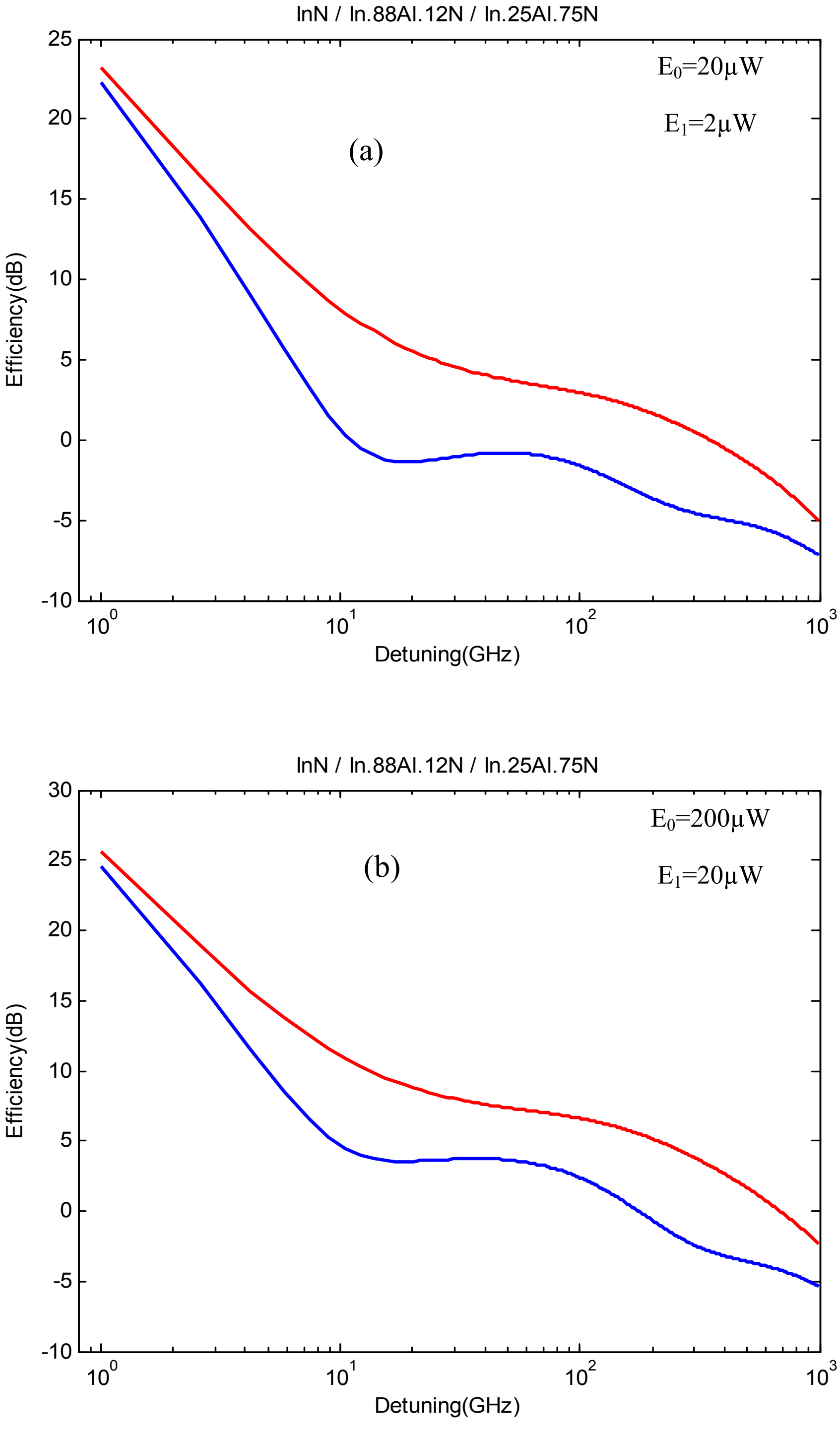

Fig. 3. The four-wave mixing efficiency $\boldsymbol{\eta}$ (dB) vs. the absolute value of the requency detuning $|\Omega|/2\pi$ in (GHz). The input power of pump and probe are (a) 20 μW and 2 μW, (b) 200 μW and 20 μW. Red for $\Omega > 0$, blue for $\Omega < 0$. At p-doped and the surface electron densities ($n_{2D}$=8.3×$10^{14}$ $m^{-2}$).

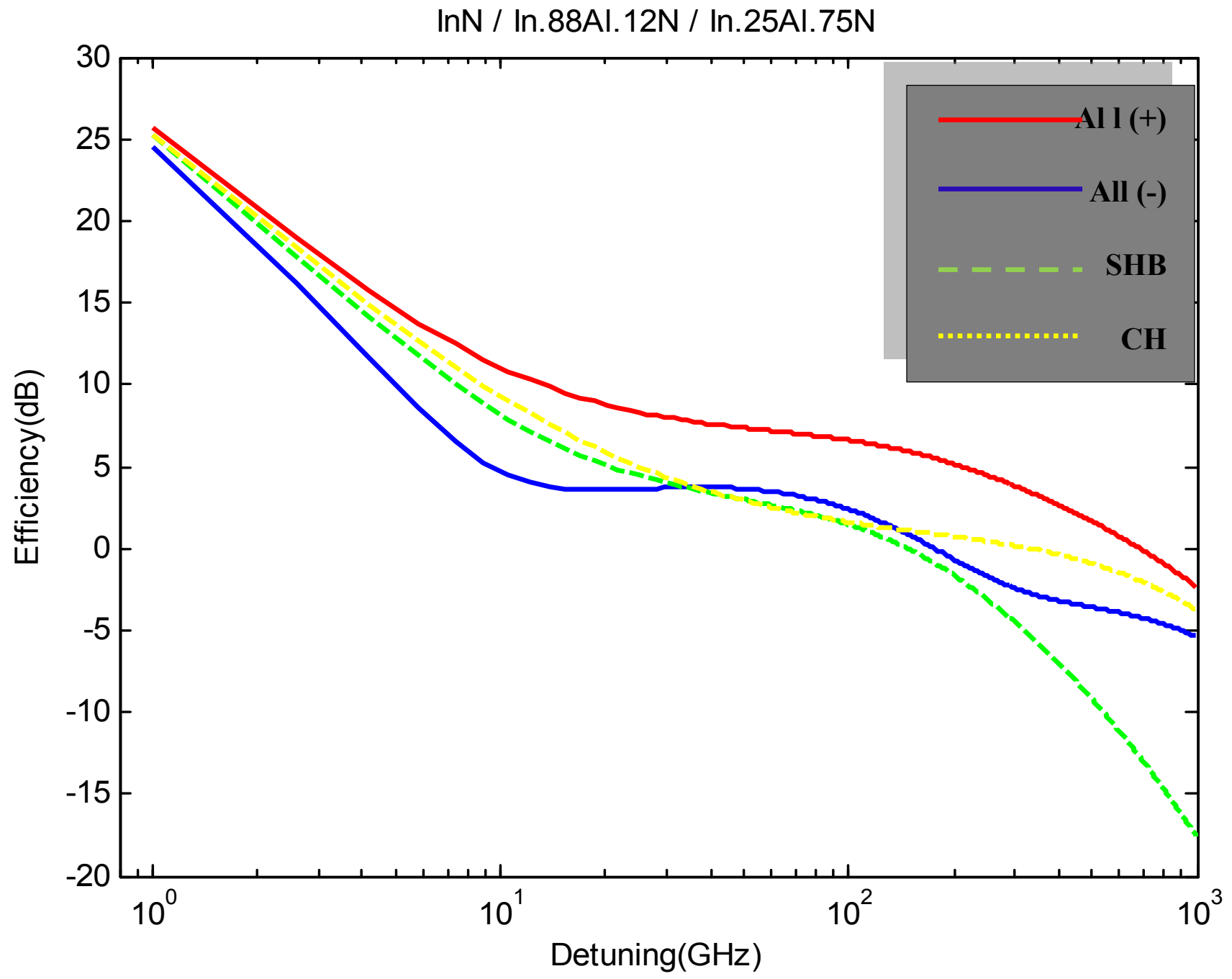

Fig. 4. Theoretical efficiency plots with individual contributions from CH and SHB superimposed for p-doped InN / $In_{088.}Al_{0.12}N$ / $In_{0.25}Al_{0.75}N$. All (−) indicates negative detuning while All (+) indicates positive detuning. The surface electron densities ($n_{2D}$=8.3×$10^{14}$ $m^{-2}$).

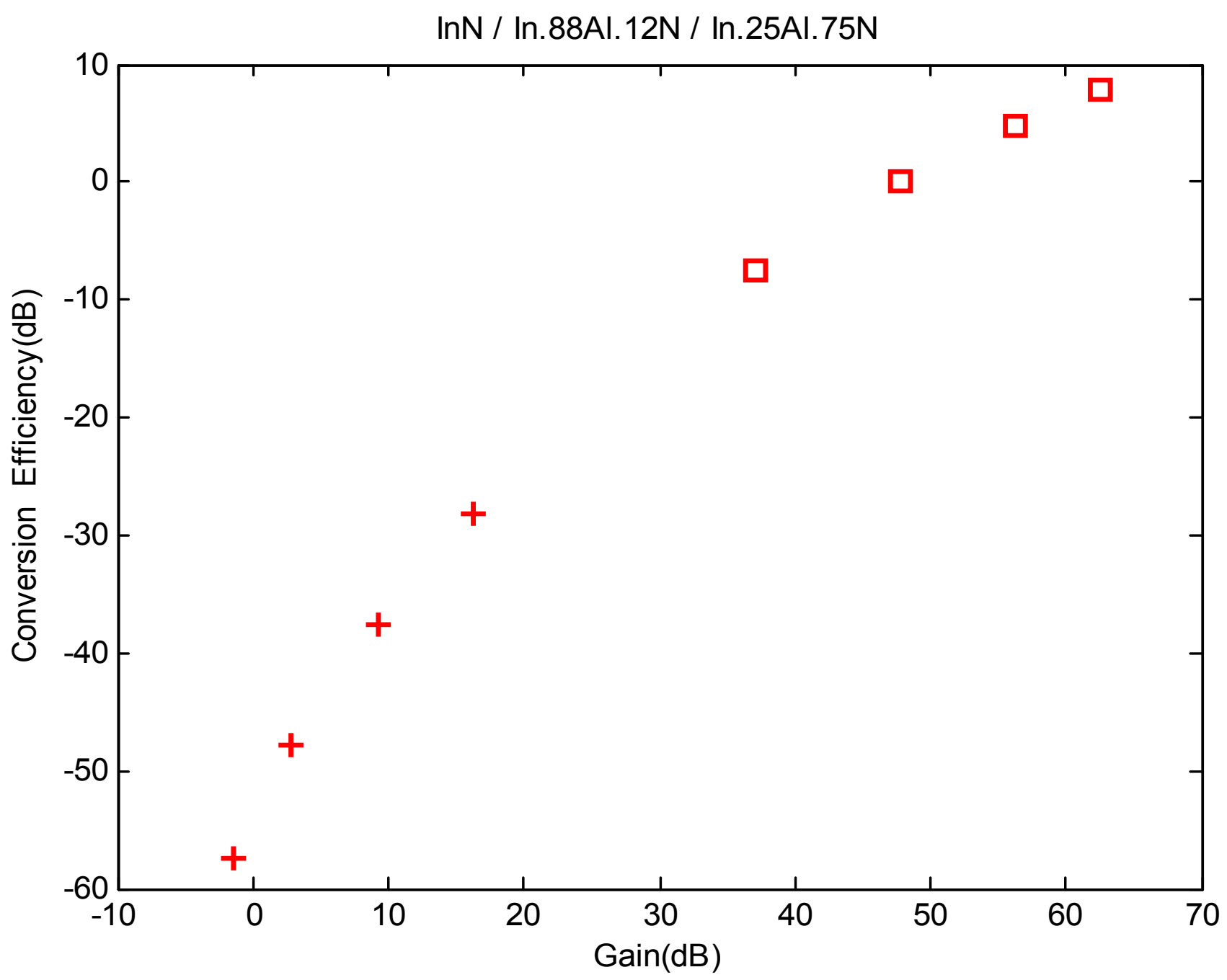

Fig. 5. Conversion efficiency (dB) vs. gain (dB), with undoped and p-doped for InN / $In_{088.}Al_{0.12}N$ / $In_{0.25}Al_{0.75}N$ structures. The input power of pump and probe is 200 μW and 20 μW, respectively, for Ω=200GHz.

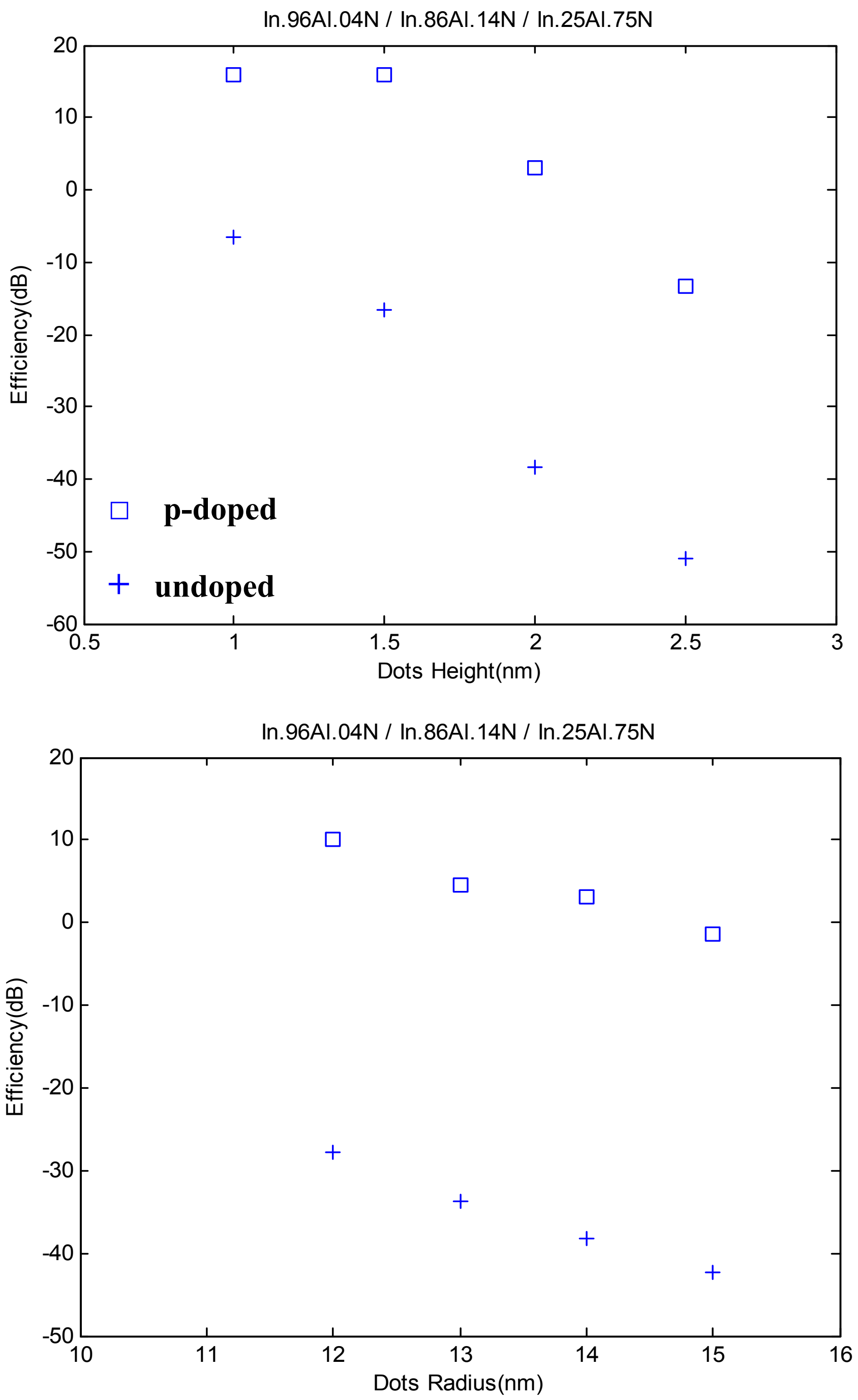

Fig. 6. Efficiency (dB) vs.: (a) dot height (nm), and (b) dot radius, with undoped and p-doped for $In_{0.96}Al_{0.04}N$ / $In_{0.86}Al_{0.14}N$ / $In_{0.25}Al_{0.75}N$ structure. The input power of pump and probe is 200 μW and 20 μW, respectively, for Ω=200GHz.

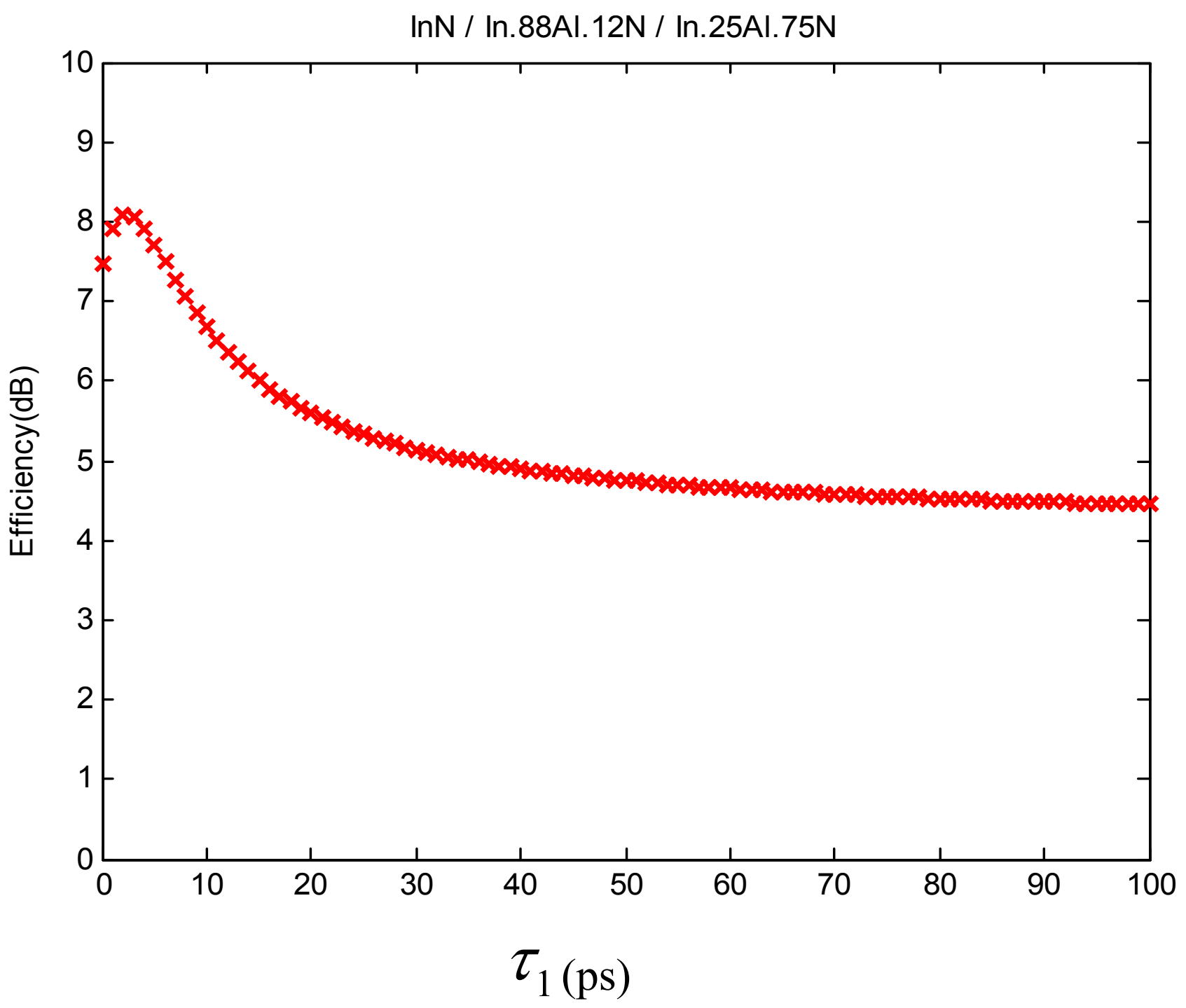

Fig. 7. Efficiency (dB) vs. $\tau_1$ (ps), for InN / $In_{088.}Al_{0.12}N$ / $In_{0.25}Al_{0.75}N$ p-doped structure. The input power of pump and probe is 200 μW and 20 μW, respectively, $\tau_2$ =0.2ps, Ω=200GHz. The surface electron density is ($n_{2D}$=8.3×$10^{14}$ $m^{-2}$).

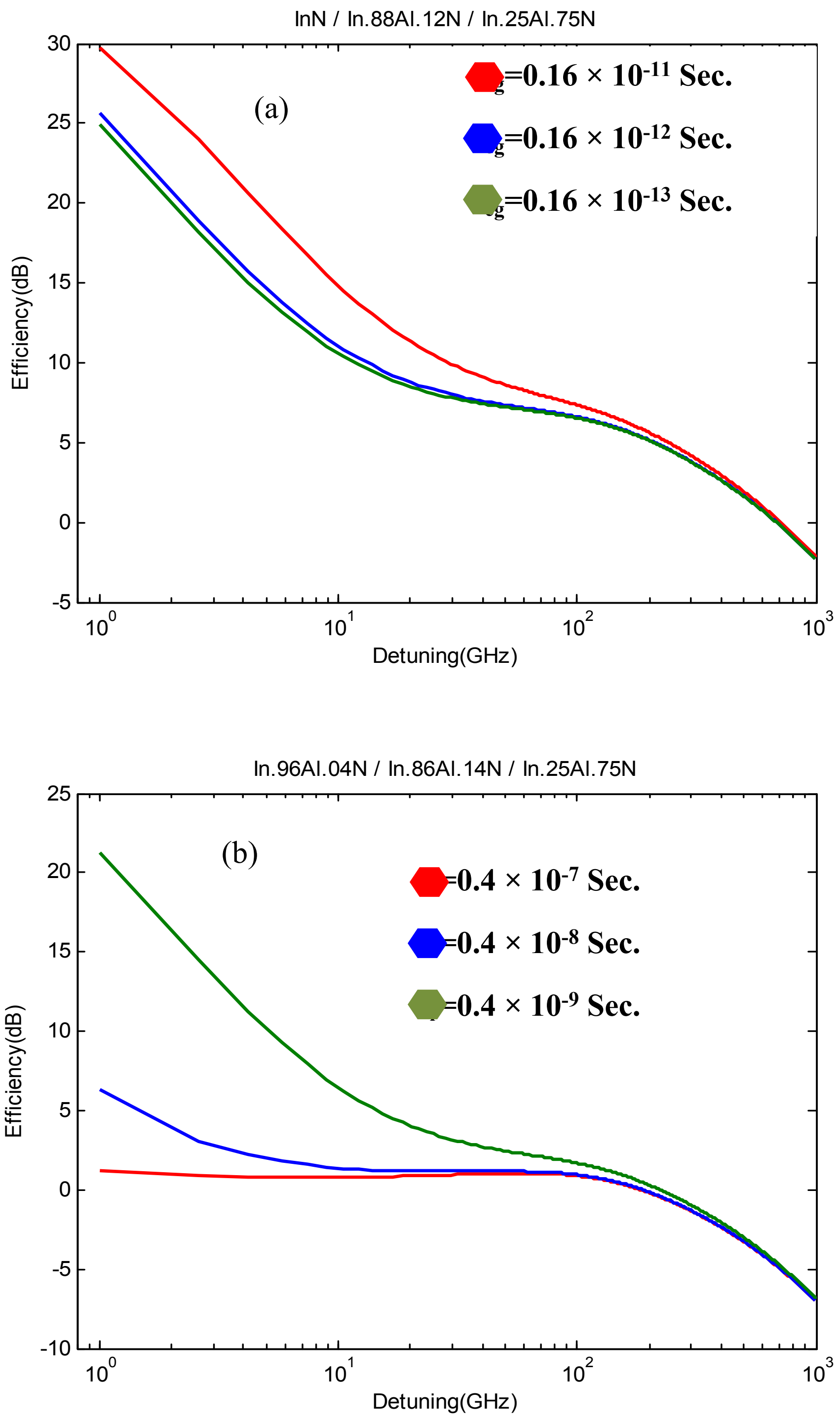

Fig. 8. The four-wave mixing efficiency $\eta$ (dB) vs. the absolute value of the frequency detuning $|\Omega|/2\pi$ in (GHz) for InAlN structure. Times (a) $\tau_{eg}$ and (b) $\tau_r$ are taken as a parameters. The input power of pump and probe is 200 μW and 20 μW. All lines for $\Omega > 0$. At p-doped and the surface electron densities ($n_{2D}=8.3\times10^{14}$ $m^{-2}$).

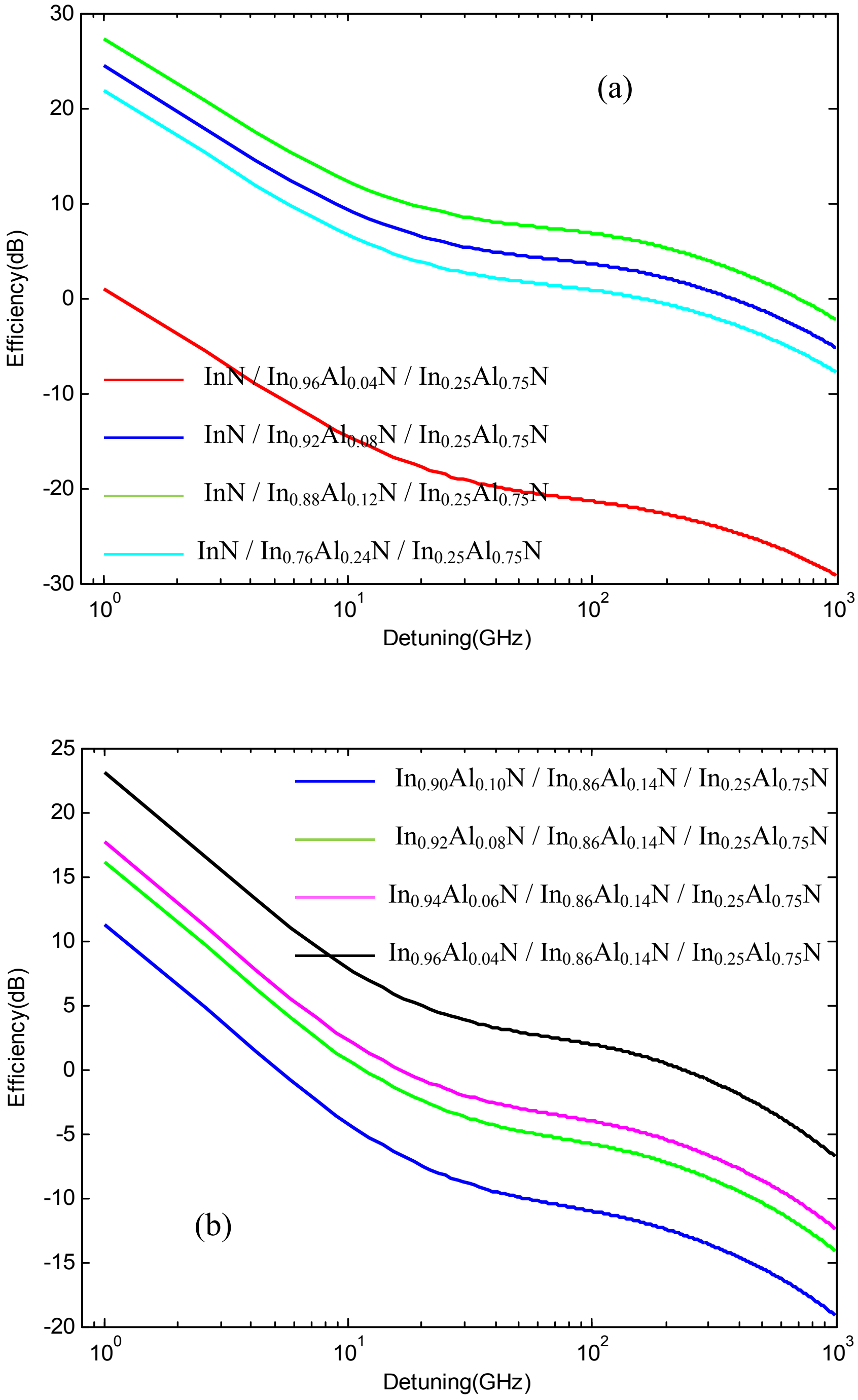

Fig. 9. The four-wave mixing efficiency $\eta$ (dB) vs. the absolute value of the frequency detuning $|\Omega|/2\pi$ in (GHz) for the branch $\Omega < 0$, for the structures (a) InN and (b) InAlN QD-SOA. The input power of pump and probe is 200 μW and 20 μW at p-doped and the surface electron densities ($n_{2D}$=8.3×$10^{14}$ $m^{-2}$).